\documentclass[conference]{IEEEtran}
\ifCLASSINFOpdf
\else
\fi
\usepackage[cmex10]{amsmath}
\usepackage{amssymb}
\usepackage{rotating} 
\usepackage{hyperref}
\hypersetup{
    colorlinks,%
    citecolor=black,%
    filecolor=black,%
    linkcolor=black,%
    urlcolor=black
}
\usepackage{float}
\usepackage{placeins}
\usepackage{stfloats}

\usepackage{url}
\makeatletter
\def\url@leostyle{%
  \@ifundefined{selectfont}{\def\UrlFont{\sf}}{\def\UrlFont{\footnotesize\ttfamily}}}
\makeatother

\makeatletter
\def\url@leotinystyle{%
  \@ifundefined{selectfont}{\def\UrlFont{\sf}}{\def\UrlFont{\footnotesize\ttfamily}}}
\makeatother

\usepackage{multirow}

\usepackage{algorithm}
\usepackage{algpseudocode}
  
\usepackage{graphicx}
\usepackage{balance}

\ifCLASSOPTIONcompsoc
\usepackage[caption=false,font=normalsize,labelfon
t=sf,textfont=sf]{subfig}
\else
\usepackage[caption=false,font=footnotesize]{subfi
g}
\fi

\newcommand{\executeiffilenewer}[3]{%
\ifnum\pdfstrcmp{\pdffilemoddate{#1}}%
{\pdffilemoddate{#2}}>0%
{\immediate\write18{#3}}\fi%
}

\usepackage{color}

\newcommand{\etal}{\textit{et al. }}

\begin{document}
\IEEEoverridecommandlockouts
\urlstyle{leotiny}

%
\title{Bypassing the Combinatorial Explosion: Using Similarity to Generate and Prioritize T-wise Test Suites for Large Software Product Lines}

\author{\IEEEauthorblockN{Christopher Henard\IEEEauthorrefmark{1},
Mike Papadakis\IEEEauthorrefmark{1},
Gilles Perrouin\IEEEauthorrefmark{2}\textsuperscript{,}\small$^\diamondsuit$\normalsize\thanks{$^\diamondsuit$FNRS Postdoctoral Researcher.}, Jacques Klein\IEEEauthorrefmark{1},
Patrick Heymans\IEEEauthorrefmark{2}\textsuperscript{,}\IEEEauthorrefmark{3} and Yves Le Traon\IEEEauthorrefmark{1}
\IEEEauthorblockA{\IEEEauthorrefmark{1}Interdisciplinary Centre for Security, Reliability and Trust (SnT), University of Luxembourg, Luxembourg, Luxembourg} 
\IEEEauthorblockA{\IEEEauthorrefmark{2}Precise Research Center In Software Engineering (PReCISE), University of Namur, Namur, Belgium}
\IEEEauthorblockA{\IEEEauthorrefmark{3}{INRIA Lille-Nord Europe, Universit\'e Lille 1 -- LIFL -- CNRS, France}\\
Email: \IEEEauthorrefmark{1}\{firstname.lastname\}@uni.lu; \IEEEauthorrefmark{2}\{firstname.lastname\}@fundp.ac.be
}}}


\maketitle

\begin{abstract}
Software Product Lines (SPLs) are families of products whose commonalities and variability can be captured by Feature Models (FMs). T-wise testing aims at finding errors triggered by all interactions amongst $t$ features, thus reducing drastically the number of products to test. T-wise testing approaches for SPLs are limited to small values of $t$ -- which miss faulty interactions -- or limited by the size of the FM. Furthermore, they neither prioritize the products to test nor provide means to finely control the generation process. This paper offers (a) a search-based approach capable of generating products for large SPLs, forming a scalable and flexible alternative to current techniques and (b) prioritization algorithms for any set of products. Experiments conducted on 124 FMs (including large FMs such as the Linux kernel) demonstrate the feasibility and the practicality of our approach.
\end{abstract}

\begin{IEEEkeywords}
SPL, Testing, T-wise Interactions, Search-based, Prioritization, Similarity
\end{IEEEkeywords}

%
\IEEEpeerreviewmaketitle

\section{Introduction}\label{intro}

A \emph{Software Product Line} (SPL) ``is a set of software-intensive systems that share a common, managed set of features satisfying the specific needs of a particular market segment or mission and that are developed from a common set of core assets in a prescribed way" \cite{clements-pl}. Features are thus the key to the discrimination of SPL members by showing their commonalities and differences. Features are often organized in a \emph{Feature Model} (FM) \cite{BenavidesSC10,Kang:1990:FODA} which represents all the possible products of the SPL by expressing relationships and constraints between features. Henceforth, we consider a \emph{product} to be a combination of features conforming to the constraints of the FM.

Testing an SPL is an inherently difficult activity \cite{springerlink:10.1007/978-3-642-14335-9_4}. Although testing all the products would be ideal, it is rarely feasible in practice. Indeed, the number of possible configurations (i.e. the products) induced by a given FM usually grows exponentially with the number of features, quickly leading to millions of possible products to test. As a result, test engineers are seeking for solutions to reduce the size of their test suites so that they can meet release deadlines and cost constraints.   

Previous work \cite{Cohen97theaetg,Kuhn04} has identified Combinatorial Interaction Testing (CIT) as a relevant approach to reduce the number of products for testing. CIT is a systematic approach for sampling large domains of test data. It is based on the observation that most of the faults are triggered by the interactions between a small number of variables. For example, Kuhn \etal  \cite{Kuhn04} have shown that 2-wise or pairwise interactions are able to disclose 80\% of the bugs. In some cases, higher interaction strengths ($t$-wise in general)  may be needed \cite{10.1109/MITP.2008.54}.  Recently, such approaches have been adapted to SPL testing \cite{Oster:2010:AIP:1885639.1885658, Perrouin2011, Johansen:2011:PRF:2050655.2050721}, generating products from the FM covering all the valid (with respect to the FM constraints) pairwise combinations of features. Some of them, like \cite{Johansen:2011:PRF:2050655.2050721}, also cover $t$-wise .

However, computing all $t$-wise interactions in the presence of constraints, as it is the case for FMs, is known to be NP-complete in the general case \cite{Johansen:2011:PRF:2050655.2050721,Perrouin:2010:AST:1828417.1828490}. As a result, although $t$-wise generation techniques from FMs have greatly improved, now relying on efficient satisfiability (SAT) solvers \cite{johansen12}, higher interaction strengths ($t>2$)  may remain inaccessible for large FMs. This is particularly problematic since 3-wise interactions were shown to commonly appear in SPL testing practice \cite{DBLP:conf/vamos/SteffensOLF12}. Since such an exact computation may remain out of reach, one may ask if it is possible to cope with these difficult situations:

[RQ1] \textit{Can we mimic $t$-wise test generation, partially but efficiently while achieving decent coverage?}         

While $t$-wise testing drastically reduces the number of products to consider, this number may still be too high to fit the budget allocated for SPL testing. 
For example, 2-wise coverage for the Linux FM (over 6,000 features) already requires 480 products to be tested \cite{johansen12}. 
Therefore, being able to prioritize the test suite with the most relevant products is critical. In this paper, the most relevant products are those that exhibit the highest number of $t$ feature interactions. The process of identifying these products is referred to as test prioritization. This forms our second research question:

[RQ2] \textit{What are the most relevant products and how to prioritize them?}

To answer RQ1, we introduce a similarity heuristic \cite{hemmati10} given in the form of a fitness function.  We assess the suitability of the function to characterize $t$-wise coverage of a given test suite.  The intuitive idea underlying this approach is: the more different the products (in terms of selected or unselected features), the more likely their ability to cover different $t$-wise interactions. We provide a search-based strategy to generate valid sets of products (i.e. respecting the constraints of the FM) for $t$-wise testing.

To answer RQ2,  we first consider that the most important products are those covering the most $t$-wise interactions. Indeed, they can potentially reveal more bugs as they test more feature interactions.  We then introduce two prioritization algorithms, named \emph{Greedy} and \emph{Near Optimal}.

Our approach introduces significant flexibility in the testing process: the number of products that can be tested (i.e. fitting the budget) can be specified as well as the time allowed for generating them. The use of similarity has the following two advantages. First, it is very fast to compute. Second, it is independent of the $t$ value. This implies that there is no need to compute and enumerate the huge number of combinations involved in the $t$ feature interactions. The applicability of the proposed strategies is evaluated on both real and generated FMs. This holds even for the largest FM, which contains over 6,000 features. The experimental data and the implementation are publicly available at \url{http://research.henard.net/SPL/}.

The remainder of this paper is organized as follows: Sections \ref{sBg} and \ref{sSim} introduce the context and the concepts underlying the proposed approaches. Sections \ref{sPrio} and \ref{sSb} respectively detail the product prioritization and generation techniques. Section \ref{sStudy} reports on the empirical study. Finally, Section \ref{sRw} discuss related work before Section \ref{sConcl} concludes the paper.


\section{Background}\label{sBg}
  
\subsection{SPL Products as Test Cases}\label{testingSPL}


In this work, we focus on a model-based testing of SPLs where the variability model is an FM. In this context, one product (an abstract test case) is represented as a set of $n$ features of an FM as $P = \{\pm f_{1}, ..., \pm f_{n}\}$, where $+f_{i}$ indicates a feature which is selected by this product, and $-f_{i}$ an unselected one. Table \ref{products_example} illustrates an example of three products and four features. For instance, product $P_{1} = \{+f_{1}, +f_{2}, +f_{3}, -f_{4}\}$ supports all the features except $f_{4}$.

\small
\begin{table}[b]

  \renewcommand{\arraystretch}{1.1}
  \caption{Example of Three Products For an FM of Four Features}
  \label{products_example}
  \centering
    \begin{tabular}{cc|c|c|c|c|l}
    \cline{3-6}
    & & \multicolumn{4}{|c|}{Features} \\ \cline{3-6}
    & & $f_{1}$ & $f_{2}$ & $f_{3}$ & $f_{4}$ \\ \cline{1-6}
    \multicolumn{1}{|c|}{\multirow{3}{*}{Products}} &
    \multicolumn{1}{|c|}{$P_{1}$} & $\times$ & $\times$ & $\times$ &  &     \\ \cline{2-6}
    \multicolumn{1}{|c|}{}                        &
    \multicolumn{1}{|c|}{$P_{2}$} & $\times$ & $\times$ &  & $\times$ &    \\ \cline{2-6}
    \multicolumn{1}{|c|}{}                        &
    \multicolumn{1}{|c|}{$P_{3}$} &  $\times$ &  & $\times$ &    \\ \cline{1-6}
    \end{tabular}
\end{table}
\normalsize

\subsection{T-wise Testing and Coverage}\label{pairwisecov}
T-wise testing focuses on the interactions between any $t \geq 2$ features of an SPL \cite{Perrouin:2010:AST:1828417.1828490}. Such an interaction is called a $t$-set. It is noted that unselected features are also involved in such interactions. For instance, with reference to Table \ref{products_example}, $(+f_{1}, -f_{2}, -f_{4})$ is a 3-set covered by $P_{3}$. The ability of a given test suite to find bugs (i.e. its fault detection power) can be estimated by the number of $t$-sets covered by the products of the test suite and is called $t$-wise coverage. In this context, $V_{t}$ denotes the set of all the valid $t$-sets of a given FM, implying that all the $t$-sets containing incompatible features, the $t$-sets where a given feature is both selected and unselected or the $t$-sets violating the constraints are excluded. More formally, $V_{t}$ can be expressed as:

\small
\begin{center}

$V_{t} = \{(\pm f_{x_{1}}, ..., \pm f_{x_{t}}) \: | \: f_{x_{1}},..., f_{x_{t}} \in  F  \wedge valid(\pm f_{x_{1}},..., \pm f_{x_{t}})\}$, 
\end{center}

\normalsize

\noindent where $F$ represents the set of features of an FM and where $valid(\pm f_{x_{1}},..., \pm f_{x_{t}})$ is a function checking the consistency of a given $t$-set with respect to the FM. Since an FM can easily be translated into a Boolean formula \cite{sat-easy}, the $valid$ function can be computed using an off-the-shelf SAT solver \cite{LP-10-1}. 

Similarly, the valid $t$-sets covered by one particular product $P$ are defined by the subset $v_{t}^{P}$, where $v_{t}^{P} \subseteq V_{t}$. In the same lines, a test suite composed of $m$ products $\{P_{1}, ..., P_{m}\}$ covers a subset of $V_{t}$, i.e. $\bigcup_{i = 1}^{m}v_{t}^{P_{i}} \subseteq V_{t}$. Thus, test suite coverage can be computed as the ratio of the number of $t$-sets covered by the test suite to the number of total valid $t$-sets:

\small
\begin{center}
$Coverage = \dfrac{\#\bigcup_{i = 1}^{m}v_{t}^{P_{i}}}{\#V_{t}}$,
\end{center}
\normalsize

\noindent where $\#A$ denotes the cardinality of the set $A$. 

Classical approaches \cite{Oster:2010:AIP:1885639.1885658, Perrouin2011, Johansen:2011:PRF:2050655.2050721} to $t$-wise testing have coverage ratio of 1 as they cover all the $t$-sets. Finally, set coverage redundancy expresses the possibility that, by removing any product, the coverage value is not altered. 


%
%
%

\section{The Similarity Heuristic}\label{sSim}
Similarity is an heuristic used here to compare two products. In model-based testing, it has been found that dissimilar test suites have a higher fault detection power than similar ones \cite{hemmati10}. The results presented in this paper (Section \ref{sStudy}) suggest that two dissimilar products are more likely to cover a greater number of valid $t$-sets than two similar ones.

In this context, we define a distance measure $d$ between two products $P_{i}$ and $P_{j}$ to evaluate their degree of similarity. As explained in Section \ref{testingSPL}, one product is considered as a set of selected or unselected features. Thus, a straightforward distance measure is a set-based one, like the Jaccard distance \cite{jaccard1901} or any other set-based distance metrics such as the Dice or Anti Dice measures \cite{hemmati10}. In our context, if $P$ represents the possible products of an SPL, the Jaccard distance is mathematically given by:

\small
\begin{displaymath}
d:
\left.
  \begin{array}{rcl}
    P \times P & \longrightarrow & [0, 1.0] \\
    (P_{i}, P_{j}) & \longmapsto & 1 - \frac{\#P_{i} \cap P_{j}}{\#P_{i} \cup P_{j}}, \;\text{where}\;P_{i}, P_{j} \in P.
  \end{array}
\right.
\end{displaymath}
\normalsize

The resulting distance varies between 0 and 1. More particularly, a distance equal to $1$ indicates that the two considered products are completely different. A distance equal to $0$ denotes that the two products are the same (redundant). It is noted that an unselected feature is also an element of the set representing a product. For instance, with reference to Table \ref{products_example},  $P_{1} = \{+f_{1}, +f_{2}, +f_{3}, -f_{4}\}$, $\textstyle P_{2} = \{+f_{1}, +f_{2}, -f_{3}, +f_{4}\}$ and $P_{3} = \{+f_{1}, -f_{2}, +f_{3}, -f_{4}\}$. Thus, $d(P_{1}, P_{2}) = 1 - \frac{\#\{+f_{1}, +f_{2}\}}{\#\{+f_{1}, +f_{2}, +f_{3}, -f_{3}, +f_{4}, -f_{4}\}} = 1 - \frac{2}{6} \approx 0.67$, $d(P_{1}, P_{3}) = 0.4$ and $d(P_{2}, P_{3})  \approx 0.86$. In this example, $P_{1}$ and $P_{3}$ are the most similar products (they share the lowest distance), whereas $P_{2}$ and $P_{3}$ are the most dissimilar ones. Thus, if we had to choose only two products, $P_{2}$ and $P_{3}$ would be the most likely to cover the greatest number of $t$-sets according to the similarity heuristic.



\section{Product Prioritization}\label{sPrio}
In this section, the similarity distances are used for prioritizing a given set of products. The aim of this process is to order a set $S$ of $m$ products $S =\{P_{1}, ..., P_{m}\}$ according to their ability to cover $t$-sets. Therefore, by testing $k\leq m$ products, the greatest possible level of coverage, for any number of $k$ products and any $t$ value, is achieved. More formally \cite{STVR:STVR430},

\emph{Given:} a set of products, $S$, the set of all the permutations of $S$, $P_{S}$ and a function $f$ from $P_{S}$ to the real numbers, $f : P_{S}  \longrightarrow \mathbb{R}_{+}$.

\emph{Problem:} finding \small$S' \in P_{S}$ \normalsize such as \small $(\forall S'' \in P_{S} | S''\neq S')[f(S')\geq f(S'')]$\normalsize. In this context, $f$ is the $t$-wise coverage achieved by $S$. To this end, two algorithms named \emph{Greedy} and \emph{Near Optimal} are introduced. They produce a list $L$, which is the result of the prioritization.

\subsection{Greedy Prioritization}
Informally, this approach iterates over the initial unordered set of products $S$, looking for the two products sharing the maximum distance. These two products are then removed from $S$ and added to the resulting list $L$. This process is repeated until all the products from $S$ are added to $L$. Algorithm \ref{greedy} formalizes this procedure.

\begin{algorithm}[b]
\caption{Greedy Prioritization$(S)$}\label{greedy}
\footnotesize
\begin{algorithmic}[1]
\State \textbf{input:} $S = \{P_{1}, ..., P_{m}\}$ \Comment{Unordered set of $m$ products}
\State \textbf{output:} $L$ \Comment{Prioritized list of $m$ products}
\State $L \gets []$ 
\While{$\#S > 0$}
\If{$\#S > 1$}
\State Select $P_{i}, P_{j}$ from $S$ \textbf{where} $max\left(d(P_{i}, P_{j})\right)$ \\ \Comment {Take the first one in case of equality}
\State $L.add(P_{i})$
\State $L.add(P_{j})$
\State $S \gets S \setminus \{P_{i}, P_{j}\}$
\Else \Comment{$S$ contains only one element}
\State $L.add(P_{i})$ \textbf{where} $P_{i}\in S$ 
\State $S \gets \emptyset$
\EndIf
\EndWhile
\State \textbf{return} $L$
\end{algorithmic}
\end{algorithm}

\subsection{Near Optimal Prioritization}
Informally, this approach selects at each step the product which is the most distant to all the products already selected during the previous steps. To this end, the two products belonging to $S$ and sharing the highest distance are first added to $L$. These two products are then removed from $S$. The next step consists in adding to $L$ and removing from $S$ the product sharing the maximum distance to all the products already added to $L$: for each product of $S$, we sum the individual distances with the other products of $L$, thus giving a value for the set.  Then the maximum is obtained by comparing these set values (Alg.\ref{no}, line 11). This process is repeated until $S$ is empty and is more formally described in Algorithm \ref{no}. 

This technique allows having more diversity than the greedy one for $k < m$ products, but it is computationally more expensive. This is due to the need of calculating all the distances from one product to the others (Alg.\ref{no}, line 11).

\begin{algorithm}[t]
\caption{Near Optimal Prioritization$(S)$}\label{no}
\footnotesize
\begin{algorithmic}[1]
\State \textbf{input:} $S = \{P_{1}, ..., P_{m}\}$ \Comment{Unordered set of $m$ products}
\State \textbf{output:} $L$ \Comment{Prioritized list of $m$ products}
\State $L \gets []$ 
\State Select $P_{i}, P_{j}$ from $S$ \textbf{where} $max\left(d(P_{i}, P_{j})\right)$ \\\Comment {Take the first ones in case of equality}
\State $L.add(P_{i})$
\State $L.add(P_{j})$
\State $S \gets S \setminus \{P_{i}, P_{j}\}$
\While{$\#S > 0$}
\State $s \gets size(L)$
\State Select $P_{i} \in S$ \textbf{where} $max\left(\sum_{j = 1}^{s}d(P_{i}, L.get(j)\right)$ \\ \Comment{Take the first one in case of equality}
\State $L.add(P_{i})$
\State $S \gets S \setminus \{P_{i}\}$
\EndWhile
\State \textbf{return} $L$
\end{algorithmic}
\end{algorithm}

\section{Search-based Product Generation}\label{sSb}
In this section, we go one step further and take benefit from the similarity heuristic to guide the generation of products (i.e. the test cases). The objective of the test generation process is to provide a set of products that fulfills the requirements of a test criterion. In the present context, this criterion is the $t$-wise coverage. If $P_{p}$ denotes the set of all the possible products and $P^{m}$ a set of $m$ products, this process is formally defined as:


\emph{Given:} a FM, the desired number of products, $m$, a given amount of time, $T$,  and a function $f$ from $P^{m}$ to the real numbers, $f : P^{m}  \longrightarrow \mathbb{R}_{+}$.

\emph{Problem:} finding $P^{m} \in P_{p}$ with respect to $T$ such as $[max(f(P_{m}))]$. In this context, $f$ is the $t$-wise coverage achieved by $S$. Toward this direction, we introduce a test generation approach, based on the (1+1) Evolutionary Algorithm \cite{DBLP:journals/tcs/DrosteJW02}. Specifically, the test generation problem is formulated as a search-based one. Instead of using complex constraints, the space of all the valid products is defined as the search space. Thus, meta-heuristic techniques can be used in order to efficiently explore this space. In view of this, similarity is used as a fitness function towards searching for products in this space. It enables: (a) a computationally interesting approach, as it will be explicitly explained in the following section and (b) prioritizing the generated products without necessitating much additional computation.

\subsection{A Similarity-based Fitness Function}\label{simfit}

Our intuition, which will be confirmed in the next section, is that the similarity heuristic is a relevant choice to define a fitness function $f$ to evaluate a set of products. Thus, if we consider a set $S$ of $m$ products $S = \{P_{1}, ..., P_{m}\}$, $f$ is formally defined as follows:

\small

\begin{displaymath}
f:
\left.
  \begin{array}{rcl}
    P^{m} & \longrightarrow & \mathbb{R}_{+} \\
    (P_{1}, ..., P_{m}) & \longmapsto & \sum_{j > i \geq 1}^{m} d(P_{i}, P_{j}).\\
  \end{array}
\right.
\end{displaymath}

\normalsize

For instance, with reference to Table \ref{products_example} and Section \ref{sSim}, $f(P_{1}, P_{2}, P_{3}) = d(P_{1}, P_{2}) + d(P_{1}, P_{3}) + d(P_{2}, P_{3}) \approx 1.93$.

This function, which generalizes the similarity distances for $m$ products, allows evaluating the quality of a set of products in terms of $t$-wise coverage. Indeed, the information conveyed by this function is: the higher the fitness value of the given set of $m$ products, the higher the distances between the products, resulting in a potentially higher $t$-wise coverage.

Although evaluating the exact coverage would be a natural choice for a fitness function, say $f_{c}$, it would be computationally expensive for such a use. Indeed, for each product, it requires computing all the $t$-sets covered by this product. Let us consider an FM with $n$ features and a set of $m$ products. If $\binom{n}{k}$ denotes the binomial coefficient, $f_{c}$ requires to compute:

\footnotesize

\begin{equation}\label{cov_binom}
 N = m\binom{n}{t} = \frac{mn!}{t!(n-t)!}
\end{equation}
\normalsize
$t$-sets to evaluate the coverage of the whole set of products, which represents $N$ operations. On the contrary, $f$ requires:
\footnotesize
\begin{equation*}
 N' = \binom{m}{2}  = \frac{m(m-1)}{2}
\end{equation*}

\normalsize

distances computation plus the sum evaluation, which represents $m$ additions.


We assume that $2 \leq t \ll n$. Therefore, the time required to compute one particular distance between two given products is small compared to the coverage evaluation of these two products, i.e. $N \gg N'$. Indeed, $f$ depends neither on $t$ nor on $n$. We also assume that one will test fewer products than the number of features, and thus that $m \ll n$. Especially, in a realistic and industrial context (with large FMs), the testing process is usually subjected to time and budget limitations. It thus does not allow testing as much products as features. It results that $N \gg N'$ and even more while $t$ increases. Recall that we focus on $t$-wise, for high $t$-values. This fact implies a computationally lower cost for $f$ compared to $f_{c}$. As a result, $f$ is used as the fitness function\footnote{Additional elements showing how the similarity distance presented in Section \ref{sSim} and how $f$ are relevant to appreciate the $t$-wise coverage of sets of products can be consulted at \url{http://research.henard.net/SPL/Resources/twise_similarity.pdf}.} for the product generation.

\subsection{Scalable Search-based Product Generation}

%
Classical constraint-based $t$-wise techniques, e.g. \cite{johansen12}, are unable to scale to large FMs and to high values of $t$. This is mainly due to the number of $t$ feature combinations. The proposed approach, which is independ of $t$, is composed of two steps. The first one is the generation of valid products using a SAT solver, and the second one is the product selection. The search process is formed by iteratively repeating these two steps. A similar technique that combines constraint solving and search-based approaches in a scalable way has been proposed by Harman \etal \cite{Harman:2011:SHO:2025113.2025144} for mutation-based test generation. 

\subsubsection{Generating Products}\label{gen_ga}

\setlength{\tabcolsep}{2.5pt}
\newcommand\colhead[1]{\begin{sideways}#1\end{sideways}} 
\begin{table*}[b]
  \renewcommand{\arraystretch}{1.1}
  \caption{120 Moderate Feature Models}
  \label{fm_small}
  \centering
{\scriptsize }%
\begin{tabular}{|l|c|c|c|c|c|c|c|c|c|c||c|c|c|c|c|c|}
\cline{2-17} 
\multicolumn{1}{l|}{} & \multicolumn{10}{c||}{
{\scriptsize Real FMs \cite{splot}}
} & \multicolumn{6}{c|}{
{\scriptsize Generated FMs}}\tabularnewline
\cline{2-17} 
\multicolumn{1}{l|}{} &

\begin{sideways}
\begin{minipage}[t]{2cm}{\begin{center}\vspace{.2em}\scriptsize Cellphone\end{center}}\end{minipage}
\end{sideways} & \begin{sideways}
\begin{minipage}[t]{2cm}{\begin{center}\vspace{.2em}\scriptsize Counter Strike Simple FM \vspace{.2em}\end{center}}\end{minipage}
\end{sideways} & \begin{sideways}

\begin{minipage}[t]{2cm}{\begin{center}\vspace{.2em}\scriptsize SPL SimulES, PnP$\;$\vspace{.2em}\end{center}}\end{minipage}
\end{sideways} & \begin{sideways}
\begin{minipage}[t]{1.8cm}{\begin{center}\vspace{.2em}\scriptsize DS Sample\vspace{.2em}\end{center}}\end{minipage}
\end{sideways} & \begin{sideways}
\begin{minipage}[t]{1.7cm}{\begin{center}\vspace{.2em}\scriptsize Electronic Drum\vspace{.1em}\end{center}}\end{minipage}
\end{sideways} & \begin{sideways}
\begin{minipage}[t]{1.8cm}{\begin{center}\vspace{.2em}\scriptsize Smart Home v2.2 \vspace{.1em}\end{center}}\end{minipage}
\end{sideways} & \begin{sideways}
\begin{minipage}[t]{1.8cm}{\begin{center}\vspace{.2em}\scriptsize Video Player\vspace{.2em}\end{center}}\end{minipage}
\end{sideways} & \begin{sideways}
\begin{minipage}[t]{2cm}{\begin{center}\vspace{.2em}\scriptsize Model Transformation\vspace{.2em}\end{center}}\end{minipage}
\end{sideways} & \begin{sideways}
\begin{minipage}[t]{1.8cm}{\begin{center}\vspace{.2em}\scriptsize Coche Ecologico\vspace{.2em}\end{center}}\end{minipage}
\end{sideways} & \begin{sideways}
\begin{minipage}[t]{2cm}{\begin{center}\vspace{.2em}\scriptsize Printers\vspace{.2em}\end{center}}\end{minipage}
\end{sideways} & 

 \begin{sideways}{\scriptsize $\;\;\;\;\;\;$ 20 FMs} \end{sideways}& \begin{sideways}{\scriptsize $\;\;\;\;\;\;$ 20 FMs} \end{sideways}& \begin{sideways}{\scriptsize $\;\;\;\;\;\;$ 20 FMs} \end{sideways}& \begin{sideways}{\scriptsize $\;\;\;\;\;\;$ 20 FMs}\end{sideways} & \begin{sideways}{\scriptsize $\;\;\;\;\;\;$ 20 FMs} \end{sideways}& \begin{sideways}{\scriptsize $\;\;\;\;\;\;$ 10 FMs}\end{sideways}\tabularnewline
\hline 
{\scriptsize $\#$Features } & {\scriptsize 11} & {\scriptsize 24} & {\scriptsize 32} & {\scriptsize 41} & {\scriptsize 52} & {\scriptsize 60} & {\scriptsize 71} & {\scriptsize 88} & {\scriptsize 94} & {\scriptsize 172} & {\scriptsize 15} & {\scriptsize 50} & {\scriptsize 100} & {\scriptsize 200} & {\scriptsize 500} & {\scriptsize 1,000}\tabularnewline
\hline 
{\scriptsize $\#$Valid products* ($\approx$)} & {\scriptsize 14} & {\scriptsize 18,176} & {\scriptsize 73,728} & {\scriptsize 6,912} & {\scriptsize 331,776} & {\scriptsize 3.87E9} & {\scriptsize 4.5E13} & {\scriptsize 1.65E13} & {\scriptsize 2.32E7} & {\scriptsize 1.14E27} & {\scriptsize 209.55} & {\scriptsize 1.02E8} & {\scriptsize 8.56E15} & {\scriptsize 3.19E20} & {\scriptsize 8.43E80} & {\scriptsize 1.81E153}\tabularnewline
\hline 
{\scriptsize $\#$Valid 2-sets} & {\scriptsize 151} & {\scriptsize 833} & {\scriptsize 1,448} & {\scriptsize 2,592} & {\scriptsize 3,746} & {\scriptsize 6,189} & {\scriptsize 7,528} & {\scriptsize 13,139} & {\scriptsize 11,075} & {\scriptsize 42,638} & {\scriptsize 300.65} & {\scriptsize 4,103.2} & {\scriptsize 17,367.8} & {\scriptsize 71,760.15} & {\scriptsize 4.67E5} & {\scriptsize 1.88E6}\tabularnewline
\hline 
\end{tabular}
\end{table*}

\begin{algorithm}[b]
\caption{Search-based Product Generation$(m,t)$}\label{sbgen}
\footnotesize
\begin{algorithmic}[1]
\State \textbf{input:} $m, t$ \Comment{Number of products to generate and execution time}
\State \textbf{output:} $L$ \Comment{List of generated products (prioritized)}
\State $L \gets []$ 
\State $S \gets \emptyset$
\For{$i \gets 1$ to $m$} 
\State $P_{\text{unpredictable}} \gets Request \;to \;the\; solver$ \\ \Comment{If the solver cannot give a new product because it has already\\\hspace{3.2em}iterated over all the valid configurations, reinitialize it}
\State $S \gets S \cup \{P_{\text{unpredictable}}\}$
\EndFor
\State $s \gets size(L)$
\While{the elapsed time is lower than $t$}
\State $fitness \gets f(L.get(1), ..., L.get(s))$
\State $L \gets \scriptstyle Greedy/Near \; Optimal \; prioritization(S)$ \Comment{Facilitated by $f$}
\State $P_{\text{worst}} \gets L.get(s)$ \Comment{\scriptsize$P_{\text{worst}}$ verifies $min\left(\sum_{k = 1}^{s}d(P_{\text{worst}}, L.get(k)\right)$\footnotesize}
\Repeat
\State $P_{\text{unpredictable}} \gets Request \;to \;the\; solver$ \\ \Comment{If the solver cannot give a new product because it has already\\\hspace{3.2em}iterated over all the valid configurations, reinitialize it}
\Until{$P_{\text{unpredictable}} \neq P_{\text{worst}}$}
\State $L.set(s, P_{\text{unpredictable}})$ \Comment{The worst product is replaced}
\State $newFitness \gets f(L.get(1), ..., L.get(s))$
\If{$newFitness \leq fitness$} \Comment{The new fitness is not better}
\State $L.set(s, P_{\text{worst}})$ \Comment{The worst product is taken back}
\EndIf
\EndWhile
\State \textbf{return} $L$
\end{algorithmic}
\end{algorithm}
A SAT solver is used to produce valid products. Once an FM is converted into a Boolean formula \cite{sat-easy}, the solver can generate valid products. As a result, a search space containing only valid products is formed.

 Typically, a product is a correct solver configuration. To this end, the literals of the logical clauses (i.e. clauses represent the constraints of the FM) are assigned values. If the constraints are satisfied, one product is returned. However, assignments to the literals are done in a particular order which involves the following problem: no uniform exploration of the space of all the valid products is possible. Indeed, the order used by the solver to parse the logical clauses and literals enables their prediction. In that case, the approach always returns the same solution in a deterministic way. As a result, the products enumeration is driven by the order used by the solver.

To overcome this issue, and thus to get products in an \emph{unpredictable} way, one solution is to randomize how the solver parses the logical clauses and the literals. It  prevents from predicting the next product that will be returned. Additionally, it allows selecting products from the full space instead of enumerating them in a predictable order.

\subsubsection{Selecting Products}


Suppose we want to generate and prioritize a list $L$ (i.e. prioritized on the fly during the search-based generation) of $m$ products. From this perspective, the search-based method informally starts by selecting $m$ products in an unpredictable way. Then, these products are evaluated by the fitness function $f$ (see Section \ref{simfit}). These products define the initial list $L$. Then, by using the distances computed while evaluating $f$, the worst product is determined. The worst product is the one which has the lowest participation in the fitness function. In other words, it is the last element of $L$. The next step consists of trying to replace this product by an unpredictable one got from the solver. This replacement is conserved if and only if the fitness of the resulting list increases. This whole process is repeated during a certain allowed amount of time $t$. It is noted that, at each iteration, the resulting list $L$ is prioritized based on the distances computed by $f$. This approach is formalized in Algorithm \ref{sbgen}. 

This technique can be considered as a genetic algorithm without crossover. It can thus be seen as an adaptation of the (1+1) Evolutionary Algorithm \cite{DBLP:journals/tcs/DrosteJW02}. Indeed, instead of removing a random product, the worst ranked product, in terms of fitness, is removed. 



\section{Empirical Study}\label{sStudy}

In this section, the product generation and prioritization approaches are assessed. In test generation, we aim at selecting products providing the highest coverage. In test prioritization, the emphasis is on maximizing the coverage increase observed each time  a product is tested. Empirical results regarding the stated research questions along with threats to the validity of this study are presented and analyzed. All the conducted experiments\footnote{The source code of the implemented approaches and the data used for the experiments are available at \url{http://research.henard.net/SPL/}.} are performed on a Quad Core@2.40 GHz with 24GB of RAM. 

The study employs 124 FMs\footnote{Handled via the SPLAR library \cite{splot} and the SAT solver Sat4j \cite{LP-10-1}.} divided into two categories. The first 120 FMs are small to medium size (with a number of features lower or equal to 1000); they are referred to as the \textit{moderate} size FMs. A second subset is composed of 4 FMs of large size; they are referred to as the \textit{large} FMs.

Regarding the moderate size FMs, 10 of them are real and 110 are artificially generated. The real FMs are taken from \cite{splot, linux-var-tools} while the artificial ones are produced with the Software Product Line Online Tools (SPLOT) FM generator \cite{splot, Mendonca:2009:SSP:1639950.1640002}. All involved FMs are consistent (i.e. the constraints are possible to fulfill). Details about the moderate FMs are recorded in Table \ref{fm_small}. It presents: the number of features, the number of valid products\footnote{Computed via a Binary Decision Diagram.} and the number of valid 2-sets.
\setlength{\tabcolsep}{.1pt}
\begin{table}[t]
  \renewcommand{\arraystretch}{1.1}
  \caption{4 Large Feature Models}
  \label{fm_large}
  \centering
{\scriptsize }%
\begin{tabular}{|l|c|c|c|c|}
\cline{2-5} 
\multicolumn{1}{l|}{} & 
\begin{minipage}{1.6cm}{\begin{center}\vspace{.2em}{\scriptsize eCos 3.0 i386pc \cite{linux-var-tools}}\vspace{.2em}\end{center}}\end{minipage}
 & 
\begin{minipage}{1.6cm}{\begin{center}\vspace{.2em}{\scriptsize FreeBSD kernel 8.0.0 \cite{linux-var-tools}}\vspace{.2em}\end{center}}\end{minipage}
 & 
\begin{minipage}{1.6cm}{\begin{center}\vspace{.2em}{\scriptsize Generated FM}\vspace{.2em}\end{center}}\end{minipage}
 & 
\begin{minipage}{1.6cm}{\begin{center}\vspace{.2em}{\scriptsize Linux kernel 2.6.28.6 \cite{linux-var-tools}}\vspace{.2em}\end{center}}\end{minipage}
\tabularnewline
\hline 
{\scriptsize $\#$Features } & {\scriptsize 1,244} & {\scriptsize 1,396} & {\scriptsize 5,000} & {\scriptsize 6,888}\tabularnewline
\hline 
{\scriptsize $\#$Valid 2-sets} & {\scriptsize 2,910,229} & {\scriptsize 3,765,597} & {\scriptsize 49,080,075} & {\scriptsize 92,540,449}\tabularnewline
\hline 
{\scriptsize $\#$Valid 3-sets ($\approx$)} & {\scriptsize 2.25E9} & {\scriptsize 3.44E9} & {\scriptsize 1.61E11} & {\scriptsize 4.19E11}\tabularnewline
\hline 
{\scriptsize $\#$Valid 4-sets ($\approx$)} & {\scriptsize 1.27E12} & {\scriptsize 2.34E12} & {\scriptsize 3.97E14} & {\scriptsize 1.50E15}\tabularnewline
\hline 
{\scriptsize $\#$Valid 5-sets ($\approx$)} & {\scriptsize 5.79E14} & {\scriptsize 1.26E15} & {\scriptsize 7.70E17} & {\scriptsize 3.85E18}\tabularnewline
\hline 
{\scriptsize $\#$Valid 6-sets ($\approx$)} & {\scriptsize 2.22E17} & {\scriptsize 5.76E17} & {\scriptsize 1.26E21} & {\scriptsize 8.71E21}\tabularnewline
\hline 
\end{tabular}
\end{table}
Concerning the large FMs, three FMs are real, taken from \cite{linux-var-tools} and one is artificially created. The details of these FMs are recorded in Table \ref{fm_large}. This table presents, for each FM, the number of features and the number of valid $t$-sets. The number of products cannot be computed in a reasonable amount of time (in days) due to the high number of constraints and features of these FMs.

For the needs of the experiment, the $t$-sets of the moderate FMs are computed using the following procedure for each FM. First, a list of all the features of the FM is recovered. Then, all the possible $t$-sets are enumerated and provided to the solver to determine whether they are valid or not. For the large FMs, computing the exact number of valid $t$-sets is a non-trivial and time consuming task. For instance, it took around 3 days to a 10-threaded program running on our system to compute the 92,540,449 valid 2-sets of the Linux FM. As $t$ increases, the number of valid $t$-sets explodes. As a result, for the large FMs and for $t\geq 3$, we estimate the number of $t$-sets. To this end, 1,000 $t$-wise sets are randomly sampled and checked. Since the total number of possible $t$-sets of an FM is known and equal to $\binom{2n}{t}$ for $n$ features ($2n$ because each feature is either selected or unselected), the valid $t$-sets can be directly estimated (law of large numbers). For example, if 800 $t$-sets out of 1,000 sampled are valid, the number of estimated valid $t$-sets is equal to $\frac{800 * \binom{2n}{t}}{1,000}$. 

\subsection{Product Generation Assessment (RQ1)}\label{prodgen}
Here, we assess the ability of the proposed approaches to cover $t$-sets. A state of the art tool named SPLCAT  \cite{Johansen:2011:PRF:2050655.2050721}, the most recent tool available handling large FMs, is employed to provide a basis for comparison.

However, SPLCAT and other approaches fail on large FMs for $2$-wise. For moderate FMs, SPLCAT does not scale well to $3$-wise or above (at least in a reasonable amount of time, in days). As a result, we can only compare our approach with SPLCAT for moderate FMs and 2-wise. As far as we know, our \textit{search-based} approach is the only one which allows scaling to any $t$ value, even for large FMs. Since no other technique can serve as a basis for comparison for the large FMs, we compare the search-based approach with products selected in an unpredictable way from the SAT solver (Section \ref{gen_ga}). In the following, this approach will be referred to as the \textit{unpredictable} one and will also serve as a comparison basis.

\subsubsection{Moderate Feature Models}\label{prodgena}
Here, we compare the search-based approach with both the unpredictable approach and SPLCAT. As a consequence, this study is only based on the 2-wise coverage and considers only the moderate FMs.
\paragraph{Experiment  Setup}
To enable a fair comparison, the search-based and unpredictable approaches generate sets of products of the same size as those provided by SPLCAT. The search-based approach is allowed to run for one minute. The results are averaged on 10 runs for each FM. 
\paragraph{Experiment Results}

\begin{figure}[t]
\centering 
\includegraphics[width=2.48in]{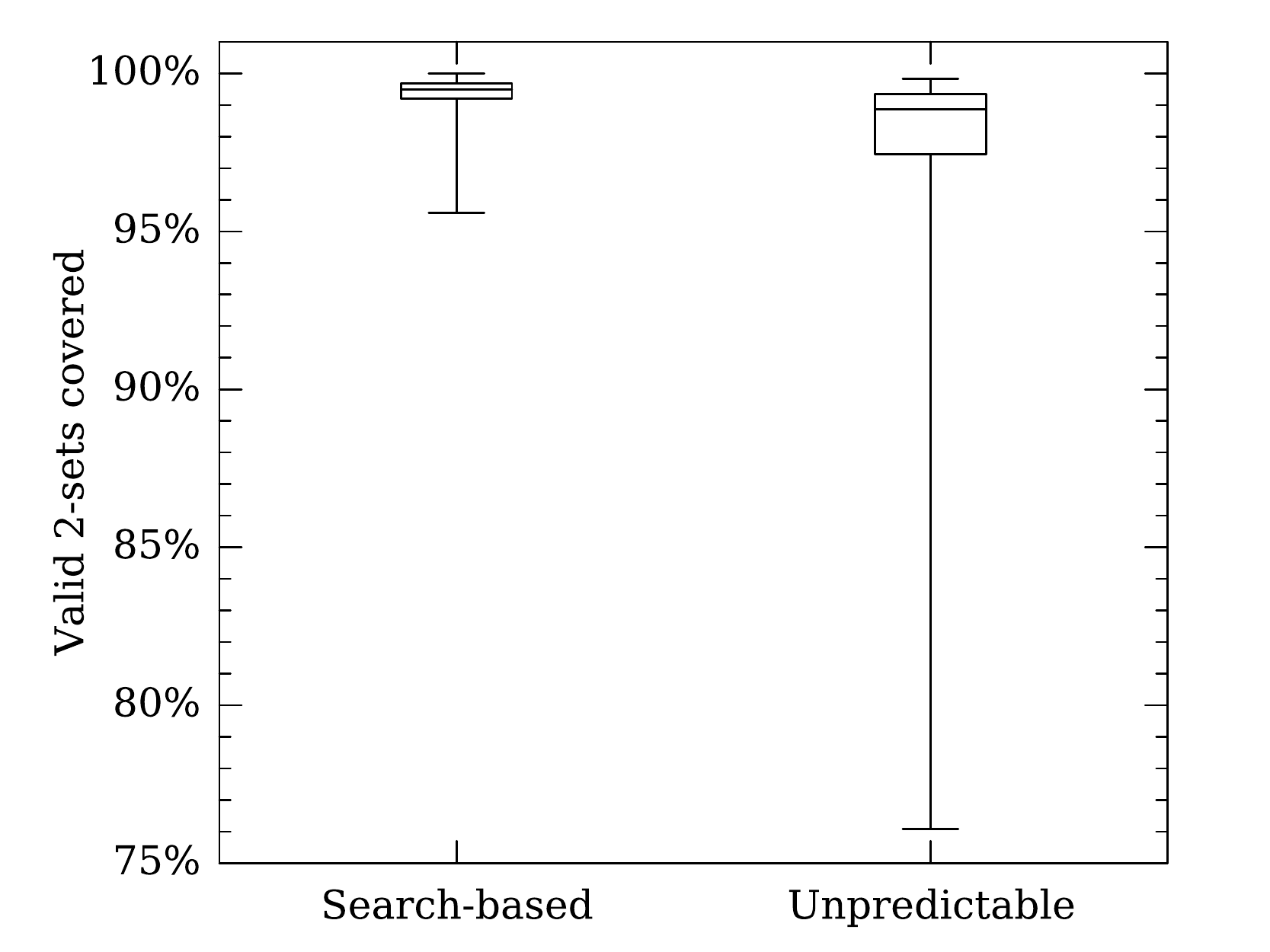}
\caption{Product Generation on the 120 Moderate FMs for $t=2$ (1 Minute Execution for the Search-based Approach, 10 Runs for Each Approach)}
\label{boxplot}

\end{figure}

The results are presented in Figure \ref{boxplot}. SPLCAT is not represented as it always achieves 100\% of coverage.  Following this figure, it appears that the proposed search-based approach, as an approximation technique, is close to SPLCAT. Indeed, in the best case, it is able to achieve 100\% of 2-wise coverage with only 1 minute of processing time allowed. In the worst case, 95\% is achieved. Besides, the search-based approach is much more stable than the unpredictable one, which can drop down to 76\% of coverage in the worst case. Although 100\% of coverage might be desirable, the focus of our approach, as explicitly stated in the introduction section, is the partial but scalable $t$-wise coverage.

Finally, the performance of SPLCAT varies. For FMs up to 200 features, SPLCAT requires less than a minute. However, it takes around 6.2 minutes for the FMs of more than 200 features, and around 159 minutes for the 1,000 features ones.

\setlength{\tabcolsep}{1.5pt}
\begin{table}[b]
  \renewcommand{\arraystretch}{1.1}
  \caption{T-wise Coverage Achieved (\%) per Approach on the Large FMs with 50 and 100 Products}
  \label{table_ga}
  \centering

{\scriptsize }%
\begin{tabular}{|c|c|cccc||cccc|}
\cline{3-10} 
\multicolumn{1}{c}{} &  & \multicolumn{2}{c|}{{\scriptsize Search-Based}} & \multicolumn{2}{c||}{{\scriptsize Unpredictable}} & \multicolumn{2}{c|}{{\scriptsize Search-Based}} & \multicolumn{2}{c|}{{\scriptsize Unpredictable}}\tabularnewline
\cline{3-10} 
\multicolumn{1}{c}{} &  & \multicolumn{4}{c||}{{\scriptsize 50 products}} & \multicolumn{4}{c|}{{\scriptsize 100 products}}\tabularnewline
\hline 
{\scriptsize FM} & {\scriptsize $t$-wise} & \multicolumn{1}{c|}{{\scriptsize Mean}} & \multicolumn{1}{c|}{{\scriptsize Std.Dev.}} & \multicolumn{1}{c|}{{\scriptsize Mean}} & {\scriptsize Std.Dev.} & \multicolumn{1}{c|}{{\scriptsize Mean}} & \multicolumn{1}{c|}{{\scriptsize Std.Dev.}} & \multicolumn{1}{c|}{{\scriptsize Mean}} & {\scriptsize Std.Dev.}\tabularnewline
\hline 
 & {\scriptsize 2} & {\scriptsize 99.12} & {\scriptsize 0.09} & {\scriptsize 98.19} & {\scriptsize 0.40} & {\scriptsize 99.62} & {\scriptsize 0.04} & {\scriptsize 99.44} & {\scriptsize 0.24}\tabularnewline
 & {\scriptsize 3} & {\scriptsize 94.53} & {\scriptsize 0.18} & {\scriptsize 92.24} & {\scriptsize 0.61} & {\scriptsize 97.55} & {\scriptsize 0.11} & {\scriptsize 96.92} & {\scriptsize 0.46}\tabularnewline
eCos & {\scriptsize 4} & {\scriptsize 83.62} & {\scriptsize 0.28} & {\scriptsize 80.85} & {\scriptsize 0.54} & {\scriptsize 91.40} & {\scriptsize 0.22} & {\scriptsize 90.51} & {\scriptsize 0.56}\tabularnewline
 & {\scriptsize 5} & {\scriptsize 67.63} & {\scriptsize 0.29} & {\scriptsize 65.64} & {\scriptsize 0.38} & {\scriptsize 80.06} & {\scriptsize 0.31} & {\scriptsize 79.56} & {\scriptsize 0.48}\tabularnewline
 & {\scriptsize 6} & {\scriptsize 50.11} & {\scriptsize 0.29} & {\scriptsize 49.36} & {\scriptsize 0.26} & {\scriptsize 64.79} & {\scriptsize 0.34} & {\scriptsize 65.05} & {\scriptsize 0.30}\tabularnewline
\hline 
 & {\scriptsize 2} & {\scriptsize 91.75} & {\scriptsize 0.12} & {\scriptsize 91.41} & {\scriptsize 0.13} & {\scriptsize 92.19} & {\scriptsize 0.12} & {\scriptsize 92.23} & {\scriptsize 0.14}\tabularnewline
 & {\scriptsize 3} & {\scriptsize 85.75} & {\scriptsize 0.18} & {\scriptsize 83.99} & {\scriptsize 0.20} & {\scriptsize 87.59} & {\scriptsize 0.16} & {\scriptsize 87.05} & {\scriptsize 0.15}\tabularnewline
{\scriptsize FreeBSD} & {\scriptsize 4} & {\scriptsize 74.94} & {\scriptsize 0.19} & {\scriptsize 71.67} & {\scriptsize 0.24} & {\scriptsize 80.82} & {\scriptsize 0.14} & {\scriptsize 79.09} & {\scriptsize 0.15}\tabularnewline
 & {\scriptsize 5} & {\scriptsize 58.54} & {\scriptsize 0.20} & {\scriptsize 55.18} & {\scriptsize 0.24} & {\scriptsize 69.74} & {\scriptsize 0.15} & {\scriptsize 67.07} & {\scriptsize 0.15}\tabularnewline
 & {\scriptsize 6} & {\scriptsize 40.39} & {\scriptsize 0.16} & {\scriptsize 37.99} & {\scriptsize 0.17} & {\scriptsize 54.30} & {\scriptsize 0.14} & {\scriptsize 51.58} & {\scriptsize 0.21}\tabularnewline
\hline 
 & {\scriptsize 2} & {\scriptsize 99.11} & {\scriptsize 0.09} & {\scriptsize 94.77} & {\scriptsize 0.20} & {\scriptsize 99.62} & {\scriptsize 0.03} & {\scriptsize 97.76} & {\scriptsize 0.19}\tabularnewline
 & {\scriptsize 3} & {\scriptsize 94.53} & {\scriptsize 0.18} & {\scriptsize 83.91} & {\scriptsize 0.22} & {\scriptsize 97.55} & {\scriptsize 0.11} & {\scriptsize 91.10} & {\scriptsize 0.27}\tabularnewline
{\scriptsize FM Generated} & {\scriptsize 4} & {\scriptsize 83.62} & {\scriptsize 0.28} & {\scriptsize 68.16} & {\scriptsize 0.20} & {\scriptsize 91.40} & {\scriptsize 0.22} & {\scriptsize 79.21} & {\scriptsize 0.25}\tabularnewline
 & {\scriptsize 5} & {\scriptsize 67.63} & {\scriptsize 0.29} & {\scriptsize 50.89} & {\scriptsize 0.17} & {\scriptsize 80.06} & {\scriptsize 0.31} & {\scriptsize 63.75} & {\scriptsize 0.23}\tabularnewline
 & {\scriptsize 6} & {\scriptsize 50.11} & {\scriptsize 0.29} & {\scriptsize 35.16} & {\scriptsize 0.18} & {\scriptsize 64.79} & {\scriptsize 0.34} & {\scriptsize 47.53} & {\scriptsize 0.26}\tabularnewline
\hline 
 & {\scriptsize 2} & {\scriptsize 96.92} & {\scriptsize 0.12} & {\scriptsize 96.05} & {\scriptsize 0.18} & {\scriptsize 97.71} & {\scriptsize 0.09} & {\scriptsize 97.28} & {\scriptsize 0.11}\tabularnewline
 & {\scriptsize 3} & {\scriptsize 91.96} & {\scriptsize 0.16} & {\scriptsize 90.49} & {\scriptsize 0.20} & {\scriptsize 94.60} & {\scriptsize 0.18} & {\scriptsize 93.82} & {\scriptsize 0.17}\tabularnewline
{\scriptsize Linux} & {\scriptsize 4} & {\scriptsize 81.37} & {\scriptsize 0.18} & {\scriptsize 79.51} & {\scriptsize 0.19} & {\scriptsize 88.53} & {\scriptsize 0.20} & {\scriptsize 87.42} & {\scriptsize 0.23}\tabularnewline
 & {\scriptsize 5} & {\scriptsize 64.42} & {\scriptsize 0.17} & {\scriptsize 62.75} & {\scriptsize 0.20} & {\scriptsize 77.38} & {\scriptsize 0.22} & {\scriptsize 76.13} & {\scriptsize 0.21}\tabularnewline
 & {\scriptsize 6} & {\scriptsize 45.24} & {\scriptsize 0.14} & {\scriptsize 44.07} & {\scriptsize 0.17} & {\scriptsize 61.13} & {\scriptsize 0.18} & {\scriptsize 60.05} & {\scriptsize 0.19}\tabularnewline
\hline 
\end{tabular}
\end{table}


%

\subsubsection{Large Feature Models}
Scaling to large FMs is a quite difficult task, even for 2-wise. Neither SPLCAT nor the tools we found (see Section \ref{sRw}) are able to “scale well to these sizes” \cite{Johansen:2011:PRF:2050655.2050721}. On the contrary, the search-based approach efficiently handles  the $t$-wise combinations where no other approach is able to do so, by producing a partial coverage. Here, we evaluate the $t$-wise coverage ability of the search-based and unpredictable approaches on the large FMs for $t= 2, ...,6$.
\paragraph{Experiment Setup}
Evaluating the $t$-wise coverage requires computing the number of unique $t$-sets covered by the products. However, this is intractable in practice due to the combinatorial explosion (each products covers $\binom{n}{t}$ $t$-sets for $n$ features). Therefore, we sample valid $t$-sets from those covered by the products. The coverage is then estimated based on the sample, considering that the latter represents all the $t$-sets covered by the products. The sampling process is repeated 10 times per each examined $t$ value ($t=2$ to $t= 6$) with samples of size 100,000. The search-based and unpredictable approaches are executed on all the large FMs to produce 5 times 50 and 100 products, with the time restriction of 30 minutes. Another experiment involves the generation of 1,000 products and the recording of the coverage over the runs of the search-based approach.
\paragraph{Experiment Results}

\setlength{\tabcolsep}{1.0pt}
\begin{table}[b]
  \renewcommand{\arraystretch}{1.1}
  \caption{6-wise Coverage and Fitness Evolution over Time for the Search-based Approach on the Large FMs with 1,000 Products}
  \label{table_ga_longruns}
  \centering

{\scriptsize }%
{\scriptsize }%
\begin{tabular}{|c|c|c|c|c|c|c|c|c|}
\cline{2-9} 
\multicolumn{1}{c|}{} & \multicolumn{2}{c|}{{\scriptsize 0 run (=unpred.)}} & \multicolumn{2}{c|}{{\scriptsize 5,000 runs}} & \multicolumn{2}{c|}{{\scriptsize 10,000 runs}} & \multicolumn{2}{c|}{{\scriptsize 15,000 runs}}\tabularnewline
\cline{2-9} 
\multicolumn{1}{c|}{} & {\scriptsize Cov.} & {\scriptsize Fit.} & {\scriptsize Cov.} & {\scriptsize Fit.} & {\scriptsize Cov.} & {\scriptsize Fit.} & {\scriptsize Cov.} & {\scriptsize Fit.}\tabularnewline
\hline 
{\scriptsize eCos} & {\scriptsize 94.191\%} & {\scriptsize 271,880} & {\scriptsize 94.225\%} & {\scriptsize 286,304} & {\scriptsize 94.263\%} & {\scriptsize 288,039} & {\scriptsize 95.343\%} & {\scriptsize 288,818}\tabularnewline
\hline 
{\scriptsize FreeBSD} & {\scriptsize 76.236\%} & {\scriptsize 294,184} & {\scriptsize 76.395\%} & {\scriptsize 299,962} & {\scriptsize 76.465\%} & {\scriptsize 300,892} & {\scriptsize 76.494\%} & {\scriptsize 301,634}\tabularnewline
\hline 
{\scriptsize FM Generated} & {\scriptsize 82.986} & {\scriptsize 258,763} & {\scriptsize 84.492\%} & {\scriptsize 263,243} & {\scriptsize 84.605\%} & {\scriptsize 263,974} & {\scriptsize 84.778\%} & {\scriptsize 264,362}\tabularnewline
\hline 
{\scriptsize Linux} & {\scriptsize 89.411\%} & {\scriptsize 296,661} & {\scriptsize 90.404\%} & {\scriptsize 298,709} & {\scriptsize 90.640\%} & {\scriptsize 299,114} & {\scriptsize 90,671\%} & {\scriptsize 299,363}\tabularnewline
\hline 
\end{tabular}
\end{table}

The results are recorded in Table \ref{table_ga}. This table presents the mean coverage achieved with respect to $t$-wise per FM and per approach. Additionally, it records the standard deviation of these values. A score above 90\% with respect to 2-wise is achieved by both the approaches and for all the studied FMs when producing 50 products. With respect to 6-wise, scores of 40\% to 50\% are achieved. By producing 100 products, higher scores are achieved for both the approaches. It should be mentioned, based on the standard deviation values recorded in Table \ref{table_ga}, that a small variation on the achieved coverage is observed. It is a fact indicating that the approaches are quite stable.


Generally, the search-based strategy provides a higher coverage compared to the unpredictable approach and especially for high values of $t$. This is true for all the $t$-wise coverage measures. Allowing more time to the search-based technique should increase the gap with the unpredictable approach since the iterations improve the set of products. However, the results are based on the selection of 50 and 100 products. Therefore, the maximum difference between the two approaches lies between the coverage of the unpredictable selection and the maximum possible coverage achievable with 50 or 100 products. Achieving 100\% of $t$-wise coverage with 50 or 100 products seems to be impossible for the large FMs. It is expected that more products are required to achieve 100\% of coverage for high values of $t$.


Table \ref{table_ga_longruns} records the coverage achieved by the search-based approach each 5,000 runs repetitions for 1,000 products with respect to 6-wise. Here, we observe that a higher level of coverage is achieved with more products. For instance, the search-based approach achieves 90,671\% of 6-wise coverage for the Linux FM. It also shows that allowing more processing time to the approach allows reaching a higher coverage. Indeed, at each 5,000 runs, the coverage recorded is higher than the previous one. Here, the unpredictable approach, represented by the ``0 run'', is also the initialization stage of the search-based strategy (Alg. 3, lines 5 to 10). For example, considering the eCos FM, 94.191\% of 6-wise coverage is achieved at the initialization. After 15,000 runs, it is 95.343\%, which represents $\approx$ 2.475744E15 additional 6-sets covered compared to the unpredictable approach. Here, it should be mentioned that the number of valid $t$-sets is extremely high (see Table \ref{fm_large}) and thus, a small increase in the coverage represents a high increase in the number of additional valid $t$-sets covered. Finally, the 15,000 runs require about 10 to 20 hours of processing time per FM.

\subsubsection{Fitness Function}

\begin{figure}[t]
\centering
\includegraphics[width=2.48in]{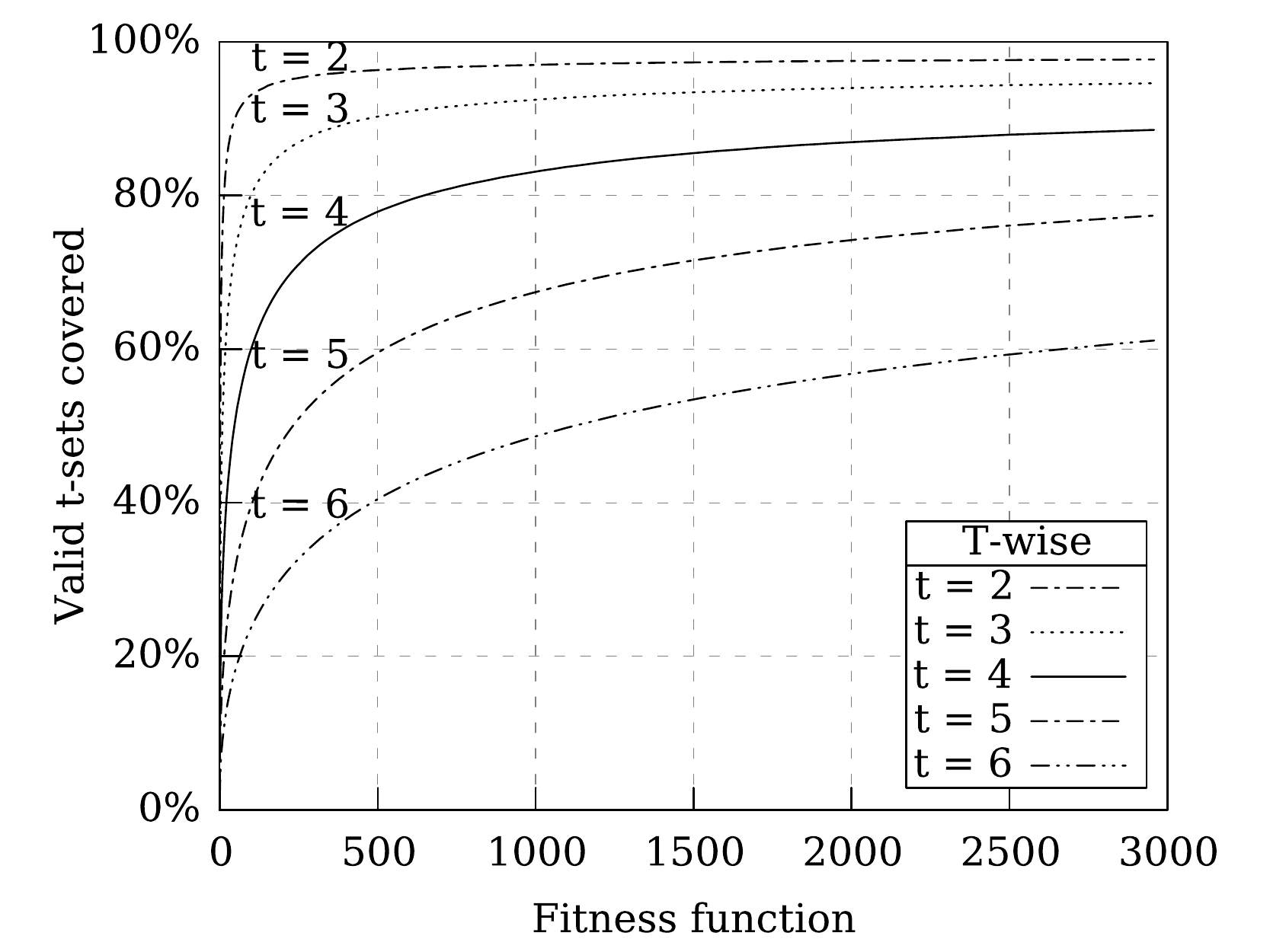}
\caption{Fitness Function Correlation with $t$-wise Coverage for the Linux FM}
\label{fitness}
\end{figure}

So far, the presented results  suggest that the search-based approach is effective and able to scale to large
FMs. Scalability is reached thanks to the ability of the similarity fitness function to mimic the $t$-wise coverage. To illustrate this
fact, Table \ref{table_ga_longruns} records the fitness function values with respect to 6-wise coverage for the large FMs as the search-based approach evolves. It shows that the fitness increases with the coverage over the runs of the approach. The same trend holds for all the FMs and values of $t$ considered in this study. Figure \ref{fitness} illustrates the correlation between the fitness and the $t$-wise coverage for the Linux FM. Therefore, the quality assessment of a set of products can be performed without computing any $t$-set, thanks to the fitness function. Recall that computing the $t$-sets  requires vast computational resources (Section \ref{simfit}, Equation \ref{cov_binom}). 


\subsubsection{RQ1}
The experiments conducted for the product generation approach emphasize the following outcomes. First, the similarity heuristic and the fitness function driving the approach form an efficient guide toward the products selection. That technique mimics $t$-wise coverage, does not depend at all on $t$ and thus avoids the combinatorial explosion due to the combinations of $t$ features. Second, the proposed technique is the first one, to the authors' knowledge, which scales well to large FMs while achieving a decent level of $t$-wise coverage (depending on the number of products desired). Finally, in addition to be a close approximation of SPLCAT, it is more flexible than the latter as it allows specifying the processing time and the number of desired products. These are characteristics conforming to an industrial context where the testing process is subjected to budget constraints.

\subsection{Similarity-based Product Prioritization Assessment}
The objective of this part is to evaluate the proposed prioritization approaches. The first experiment focuses on $t=2$ for the moderate FMs, due to the limitations of SPLCAT (see Section \ref{prodgen} and \ref{prodgena}). The second experiment demonstrates its ability to scale to any $t$ value for the large FMs.

To compare the prioritization approaches, the area under curve is evaluated \cite{Do:2006:UMF:1248725.1248767}. This area is the  numerical approximation of the integral of the coverage curve and is computed using the trapezoidal rule, i.e. $\int_a^b g(x)dx \approx (b-a) \frac{g(a)+g(b)}{2}$. Thus, for each prioritization method, if $cov(x)$ denotes the percentage of $t$-wise coverage achieved with the $x$-th product, then the area value is given by $\sum_{i = 1}^{99} \int_{i}^{i+1} cov(x)dx = \sum_{i = 1}^{99} \frac{cov(i) + cov(i+1)}{2}$. A higher area under curve value expresses a more effective prioritization.

\subsubsection{Moderate Feature Models}
This experiment focuses on comparing our prioritization techniques with SPLCAT for $t=2$. This tool does not prioritize the products, but it tries to cover the maximum of $2$-sets each time a product is added. The resulting products can thus be considered as ordered for covering faster the highest amount of 2-sets. 
\paragraph{Experiment  Setup}

 For each moderate FMs, three different sets of products are used to apply the prioritization techniques. The first set is the set of products produced by SPLCAT (Case I). The second one is a set of $n$ products, where $n=\frac{\#features}{2}$, selected with the unpredictable method (Case II). Finally, the last set is composed of the products generated by SPLCAT plus the same amount of products selected by the unpredictable method (Case III). Using these different sets allows ensuring that the prioritization approaches are relevant whatever the nature of the products. 
\begin{table*}[t]
  \renewcommand{\arraystretch}{1.1}
  \caption{Prioritization Results: Area Under Curve (scale 1:1,000)}
  \label{area_prio}
  \centering

{\scriptsize }%
\begin{tabular}{|c|c|c|c||c|c|c|c|c|c|c|c|c|c|c|c|c|c|c|c|c|c|c|c|}
\cline{2-24} 
\multicolumn{1}{c|}{} & {\scriptsize Case I} & {\scriptsize Case II} & {\scriptsize Case III} & \multicolumn{5}{c|}{{\scriptsize Case IV with 100 products }} & \multicolumn{5}{c|}{{\scriptsize Case IV with 500 products}} & \multicolumn{5}{c|}{{\scriptsize Case V with 100 products}} & \multicolumn{5}{c|}{{\scriptsize Case V with 500 products}}\tabularnewline
\hline 
{\scriptsize Technique \textbackslash{} T-wise} & {\scriptsize 2} & {\scriptsize 2} & {\scriptsize 2} & {\scriptsize 2} & {\scriptsize 3} & {\scriptsize 4} & {\scriptsize 5} & {\scriptsize 6} & {\scriptsize 2} & {\scriptsize 3} & {\scriptsize 4} & {\scriptsize 5} & {\scriptsize 6} & {\scriptsize 2} & {\scriptsize 3} & {\scriptsize 4} & {\scriptsize 5} & {\scriptsize 6} & {\scriptsize 2} & {\scriptsize 3} & {\scriptsize 4} & {\scriptsize 5} & {\scriptsize 6}\tabularnewline
\hline 
{\scriptsize Random} & {\scriptsize 8.51} & {\scriptsize 8.37} & {\scriptsize 9.24} & {\scriptsize 9.23} & {\scriptsize 8.34} & {\scriptsize 7.10} & {\scriptsize 5.57} & {\scriptsize 4.05} & {\scriptsize 49.06} & {\scriptsize 47.69} & {\scriptsize 45.09} & {\scriptsize 40.88} & {\scriptsize 35.22} & {\scriptsize 8.65} & {\scriptsize 7.33} & {\scriptsize 5.67} & {\scriptsize 4.02} & {\scriptsize 2.67} & {\scriptsize 47.47} & {\scriptsize 44.70} & {\scriptsize 40.42} & {\scriptsize 34.40} & {\scriptsize 27.40}\tabularnewline
\hline 
{\scriptsize Greedy} & {\scriptsize 8.84} & {\scriptsize 8.50} & {\scriptsize 9.35} & {\scriptsize 9.28} & {\scriptsize 8.43} & {\scriptsize 7.17} & {\scriptsize 5.61} & {\scriptsize 4.07} & {\scriptsize 49.16} & {\scriptsize 47.77} & {\scriptsize 45.15} & {\scriptsize 40.97} & {\scriptsize 35.32} & {\scriptsize 8.89} & {\scriptsize 7.68} & {\scriptsize 6.13} & {\scriptsize 4.45} & {\scriptsize 3.03} & {\scriptsize 47.76} & {\scriptsize 45.28} & {\scriptsize 41.44} & {\scriptsize 36.14} & {\scriptsize 29.55}\tabularnewline
\hline 
{\scriptsize Near Optimal} & {\scriptsize 8.93} & {\scriptsize 8.60} & {\scriptsize 9.44} & {\scriptsize 9.33} & {\scriptsize 8.48} & {\scriptsize 7.22} & {\scriptsize 5.65} & {\scriptsize 4.11} & {\scriptsize 49.23} & {\scriptsize 47.92} & {\scriptsize 45.46} & {\scriptsize 41.32} & {\scriptsize 35.66} & {\scriptsize 9.06} & {\scriptsize 8.00} & {\scriptsize 6.56} & {\scriptsize 4.91} & {\scriptsize 3.37} & {\scriptsize 48.15} & {\scriptsize 46.19} & {\scriptsize 42.95} & {\scriptsize 38.03} & {\scriptsize 31.71}\tabularnewline
\hline 
{\scriptsize SPLCAT} & {\scriptsize 8.88} & \multicolumn{1}{c}{} & \multicolumn{1}{c}{} & \multicolumn{1}{c}{} & \multicolumn{1}{c}{} & \multicolumn{1}{c}{} & \multicolumn{1}{c}{} & \multicolumn{1}{c}{} & \multicolumn{1}{c}{} & \multicolumn{1}{c}{} & \multicolumn{1}{c}{} & \multicolumn{1}{c}{} & \multicolumn{1}{c}{} & \multicolumn{1}{c}{} & \multicolumn{1}{c}{} & \multicolumn{1}{c}{} & \multicolumn{1}{c}{} & \multicolumn{1}{c}{} & \multicolumn{1}{c}{} & \multicolumn{1}{c}{} & \multicolumn{1}{c}{} & \multicolumn{1}{c}{} & \multicolumn{1}{c}{}\tabularnewline
\cline{1-2} 
\end{tabular}
\end{table*}
\begin{figure*}[b]
\centerline{
\subfloat[Case I: SPLCAT products]{\includegraphics[width=2.45in]{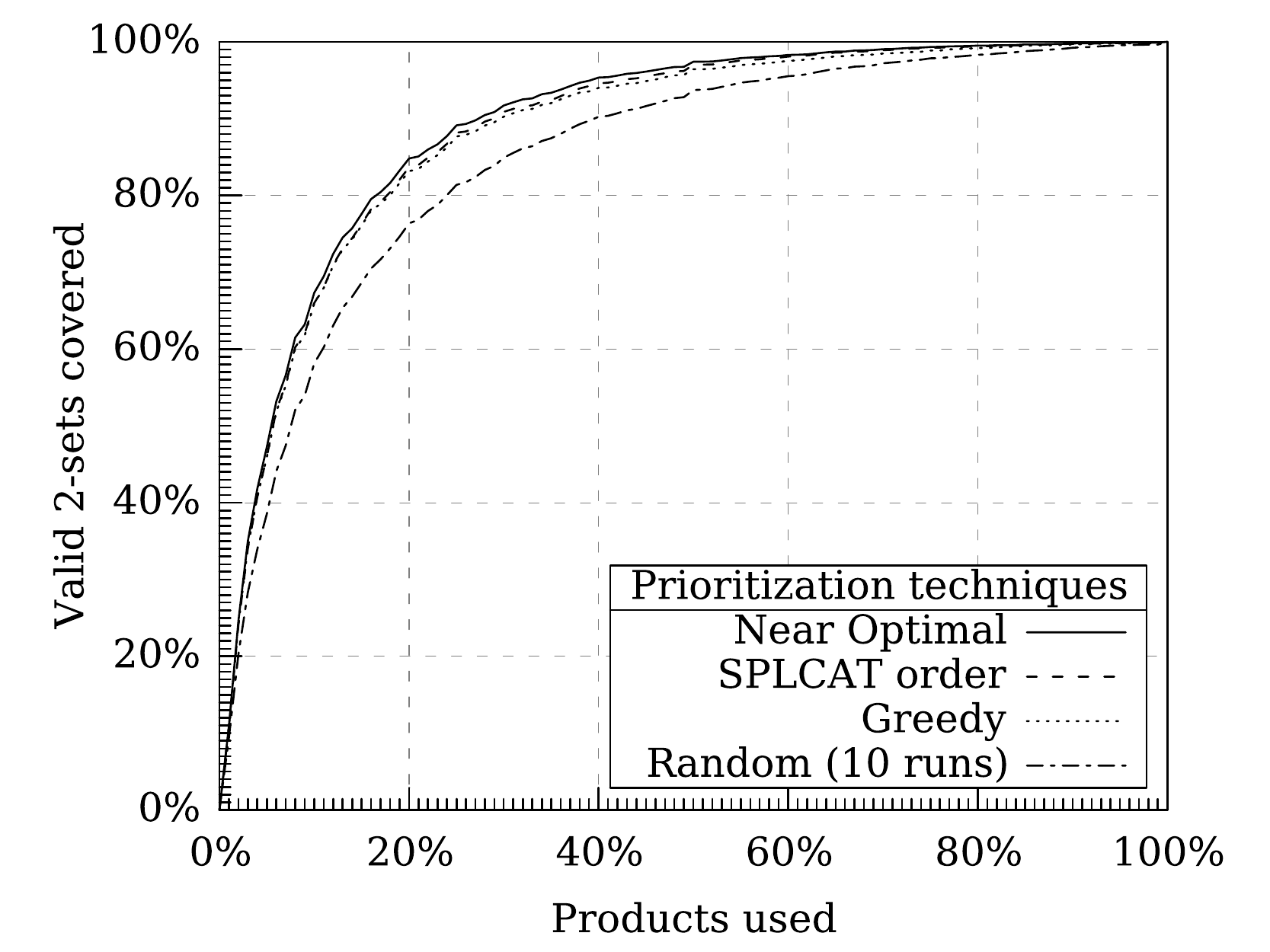}
\label{prio_1_case}}
\hfil
\subfloat[Case II: unpredictable products]{\includegraphics[width=2.45in]{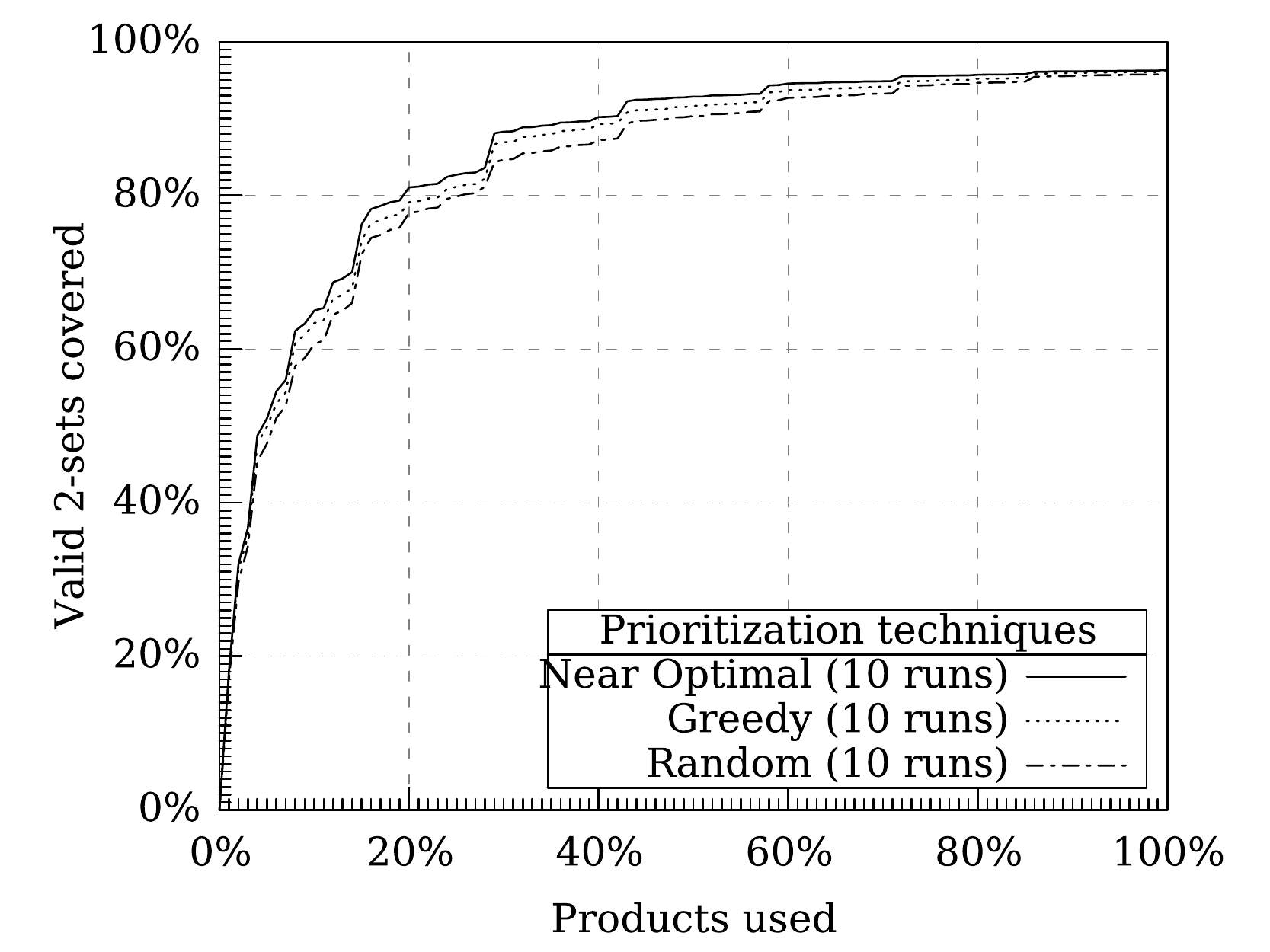}
\label{prio_3_case}}
\hfil
\subfloat[Case III: SPLCAT + unpredictable products]{\includegraphics[width=2.45in]{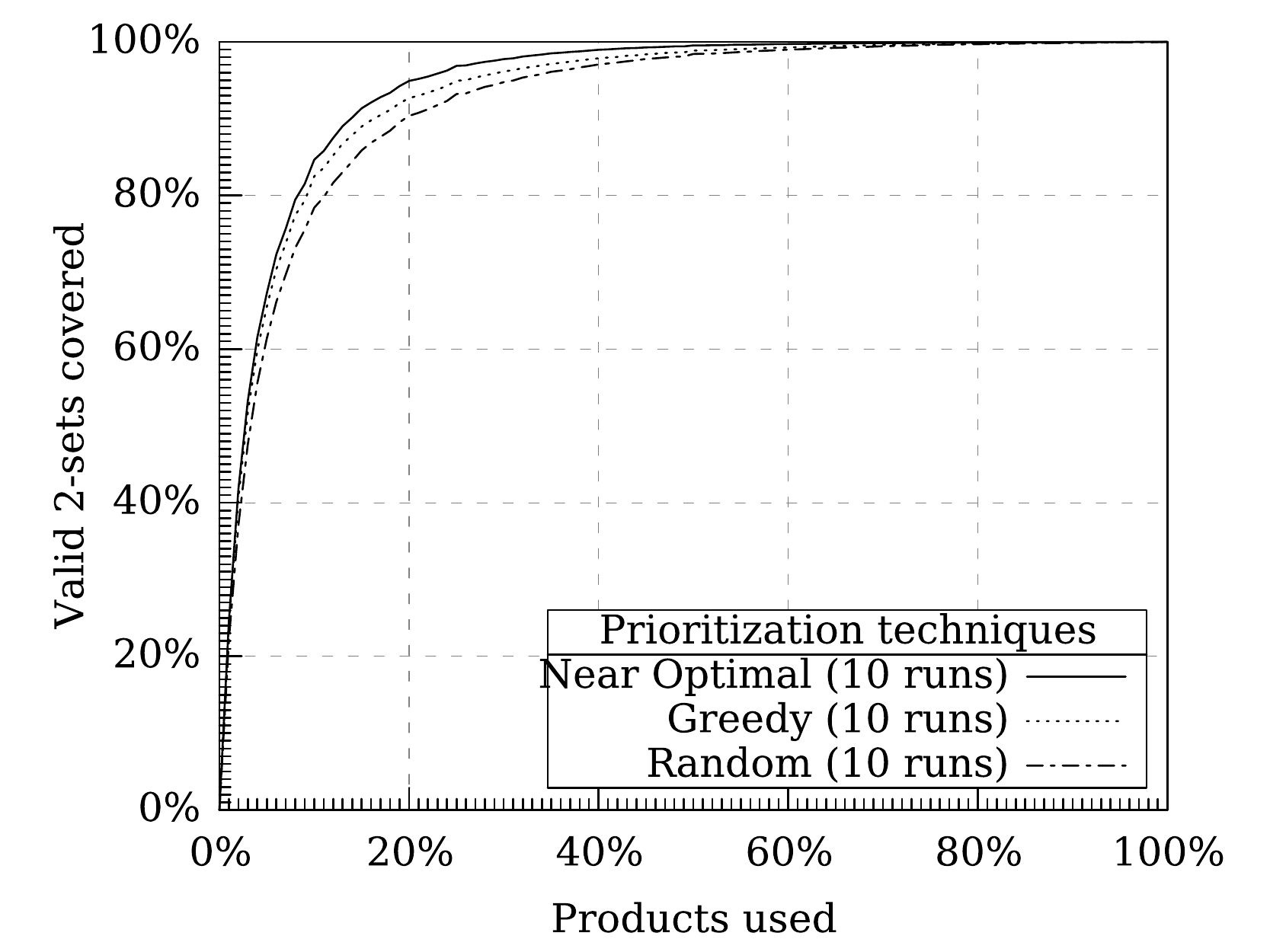}
\label{prio_5_case}}}
\caption{Prioritization on Moderate FMs ($t=2$)}
\label{fig_prio}
\end{figure*}

All these sets of products are randomized before executing the prioritization techniques. This practice ensures that our approaches are independent of the original order. On each of the three cases and for each FM, a random prioritization is averaged 10 times. Cases II and III are independently repeated 10 times to avoid any bias from the initial set of products.
\paragraph{Experiment Results}\label{res_prio_mod}

\setlength{\tabcolsep}{2.3pt}


Table \ref{area_prio} presents the area under curve for each case and technique. Recall that a higher surface value indicates a better prioritization. With respect to Table \ref{area_prio} and focusing on Case I, we observe the following ordering: Random $<$ Greedy $<$ SPLCAT $<$ Near Optimal. For Case II and Case III, the order Random $<$ Greedy $<$ Near Optimal is observed. Thus, in all the cases, the Near Optimal prioritization provides the best prioritization as its area under curve is the greatest one. 

Figure \ref{fig_prio} illustrates this behavior. For each case, the results are averaged on all the FMs by normalizing the number of products selected from 0 to 100\%. For instance, with respect to Case I, the Near Optimal prioritization approach enables covering more than 90\% of the 2-set with only 28\% of the products. On the contrary, the random prioritization needs more than 40\% of the products.  For Case II and Case III (Figure \ref{prio_3_case} and \ref{prio_5_case}), the same trends are observed. These results emphasize that the prioritization techniques are either able to perform similarly (Greedy) or better (Near Optimal) as SPLCAT. 


\subsubsection{Large Feature Models}
This experiment focuses on the evaluation of the Near Optimal and Greedy prioritization on the large FMs and for $t = 2 $ to $t=6$.
\paragraph{Experiment Setup}
We generate two sets of 100 and 500 products containing dissimilar products (Case IV) and two sets of the same sizes containing half similar and dissimilar products (Case V). We choose these two kinds of sets of products since the prioritization approaches are similarity-driven and can thus be influenced by the nature of the used sets. Indeed, applying these approaches on sets containing dissimilar products can be less effective than applying them on sets containing similar products. We randomize each set of products and execute the Greedy and Near Optimal prioritizations on each of them. We also produce 10 random orderings to compare with our approaches. This practice shows that the prioritization techniques are not affected by random orders.

\paragraph{Experiment Results}
As for the results presented in Section \ref{res_prio_mod}, we evaluate the area under curve. The results are recorded in Table \ref{area_prio}. The random is averaged on 10 runs for each value of $t$. The presented values are averaged on the 4 large FMs. We observe the following ordering for both Case IV and Case V: Random $<$ Greedy $<$ Near Optimal. This shows that the prioritizations approaches are relevant for finding the dissimilarities in the sets containing both similar and dissimilar products. The Near Optimal prioritization tends to be the most relevant approach.
%

\begin{figure}[t]
\centerline{
\subfloat[Dissimilar Set of Products]{\includegraphics[width=2.48in]{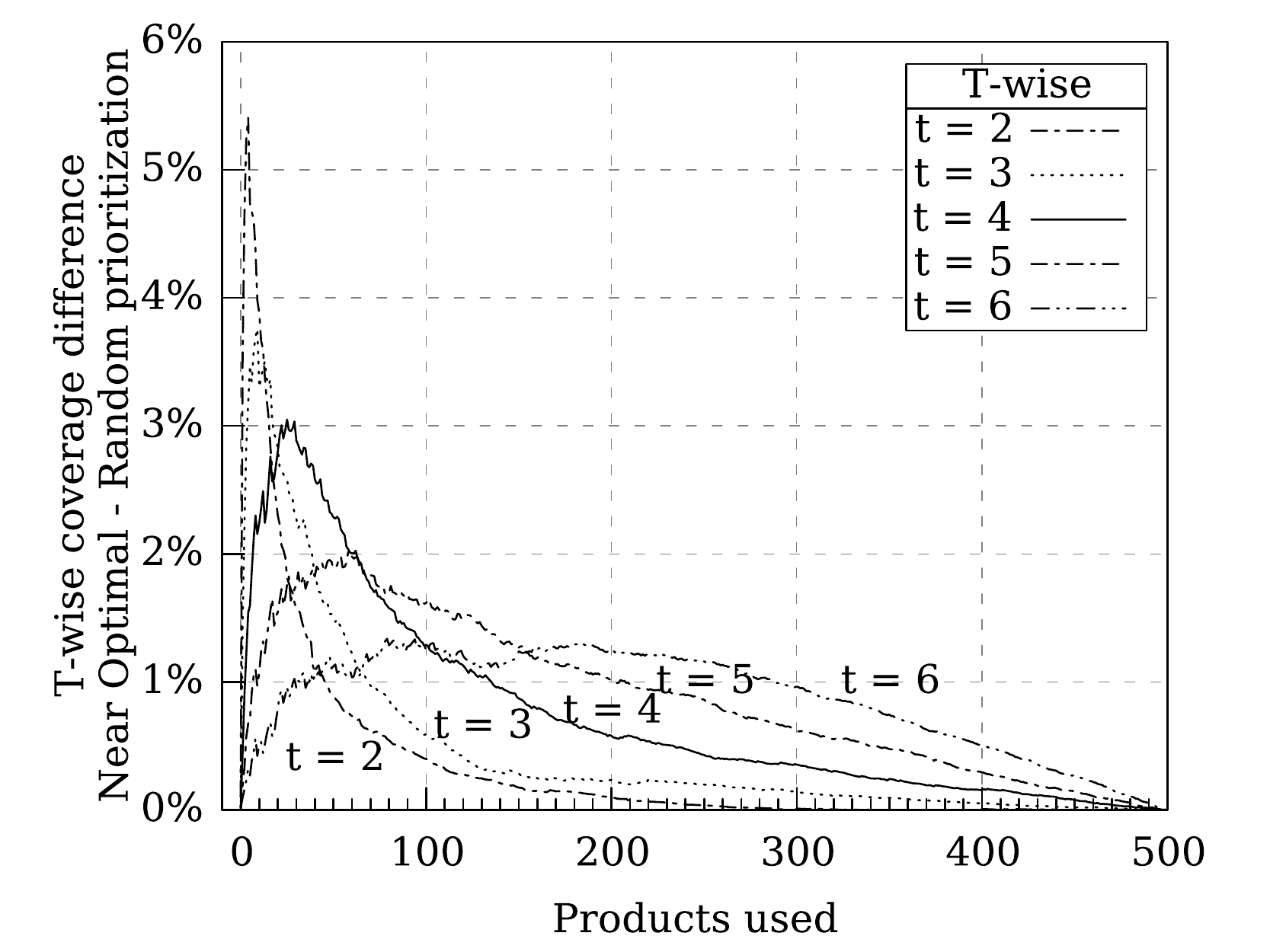}
\label{priotdis}}}
\centerline{
\subfloat[Similar + Dissimilar Products]{\includegraphics[width=2.48in]{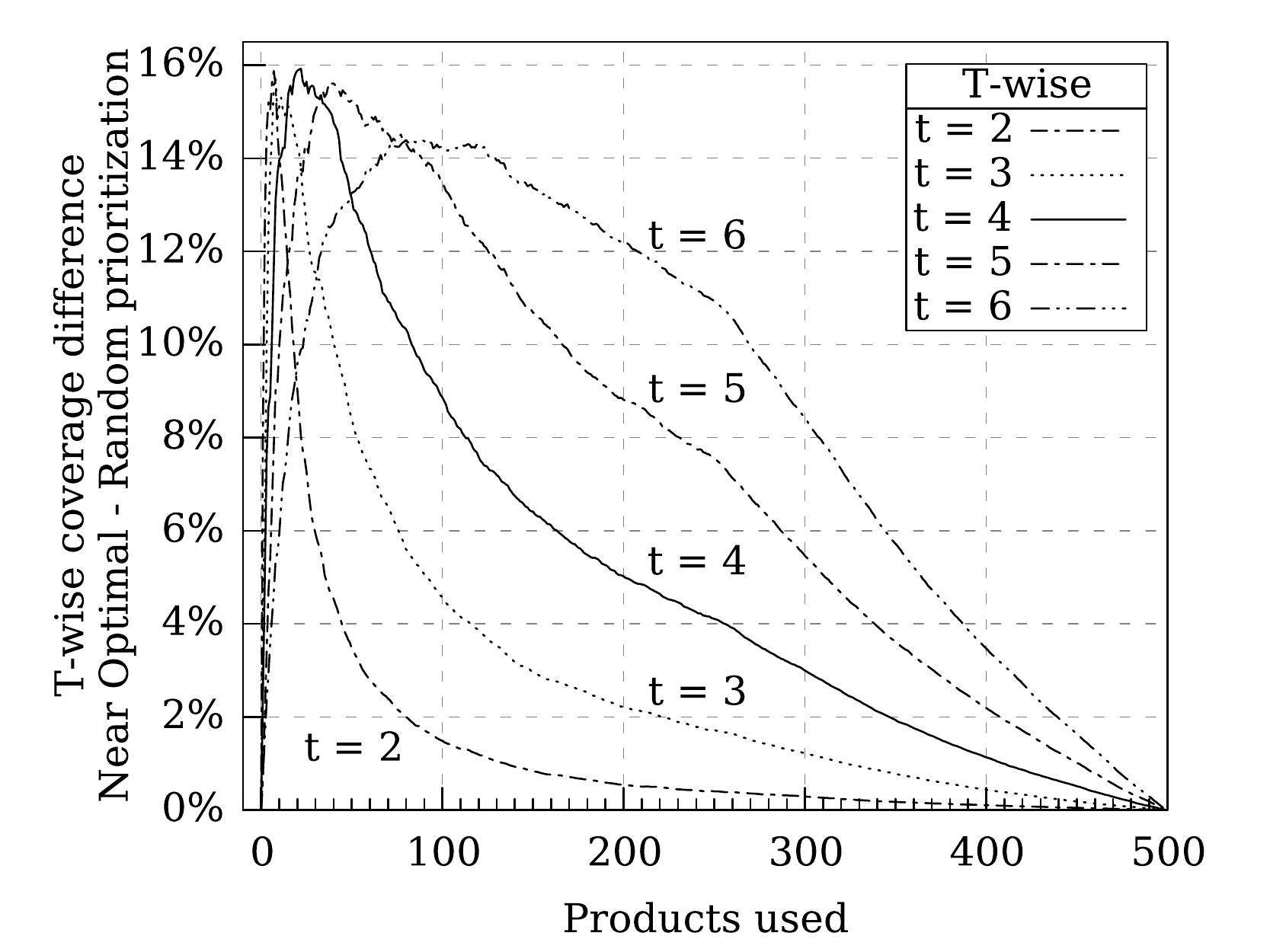}
\label{priotsim}}}
\caption{Near Optimal VS Random Prioritization on Large FMs}
\label{priot}

\end{figure}

As expected, when products are already dissimilar (Case IV), the gain is lesser than when the set of products is any (Case V). Additionally, Figure \ref{priot} presents the $t$-wise coverage difference between the Near Optimal prioritization and the random ordering for 500 products, averaged on the 4 FMs. For Case IV (Figure \ref{priotdis}) and $t=4$, 3\% of difference is observed with 30 products selected. For Case V (Figure \ref{priotsim}), 14\% of difference is observed with 100 products for $t=6$. 

\subsubsection{RQ2}

The experiments conducted for the prioritization bring out the following observations. First, our approaches compete with existing ones that could produce a kind of ordering, like SPLCAT. Second, the most relevant products contributing to $t$-wise coverage are the most dissimilar ones. This is enabled by the similarity heuristic. Finally, the proposed prioritization approaches are able to prioritize any set of products, by looking for the dissimilarities. This is performed without computing any $t$-sets and regardless of the value of $t$.

\subsection{Threats to Validity}
Although we used various FMs, there is an \emph{external validity} threat. Indeed, we cannot ensure that the proposed strategies will provide similar results on different sets of FMs (larger or more constrained). To reduce this threat, we used a relatively large set of 124 FMs of different sizes, combined real and generated FMs to cope with a variety of situations. Additionally, potential errors in our implementation could affect the presented results and lead to \emph{internal validity} threats. To overcome these threats, we divided the implementation into sub stages to have a better control on each of the steps composing the proposed approaches. The comparison with SPLCAT also gave us confidence in our implementation. Besides that, to prevent as possible a \emph{construct validity} threat, we sampled each technique on 10 runs. To enable reproducibility and to reduce the above-mentioned threats, we made our implementation and the experiment data publicly available.

\section{Related Work}\label{sRw}
\subsection{Prioritization}

As surveyed by Nie and Leung  \cite{Nie:2011:SCT:1883612.1883618}, efforts have been made to prioritize test suites.  For instance, Bryce and Colbourn \cite{bryce2007one} use search-based techniques (e.g. hill climbing) to select the ``best test'' in terms of $t$-wise coverage.  We share their motivation of focusing on the most relevant test.  Additionally, the proposed techniques offer improvements over the ``natural'' ordering provided by the AETG algorithm \cite{Cohen97theaetg} in line with our experimentations. However, computing $t$-wise coverage for each product is expensive, especially for constrained cases, which are not taken into account in their approach and thus unsuitable in SPLs context. Some approaches combine $t$-wise prioritization and generation (see below).  Finally, there are SPL-dedicated efforts, also in the context of test generation, but not directed to $t$-wise, such as Uzuncaova \etal \cite{uzuncaova-2008} work.          

\subsection{T-wise generation}

SPL $t$-wise testing approaches typically fall into two categories:  \emph{constraint-based} and \emph{search-based}.  

\subsubsection{Constraint-based Approaches}

Since $t$-wise testing of SPLs is made difficult by the presence of constraints \cite{Cohen2007}, the use of constraint solving solutions have been investigated. In Perrouin \etal work \cite{Perrouin:2010:AST:1828417.1828490}, a solution based on Alloy, a SAT solver, was devised.  The approach was non-predictable in terms of generated solutions and strategies to improve scalability were proposed. Oster \etal \cite{Oster:2010:AIP:1885639.1885658} optimized the problem upfront by flattening the FM and using CIT algorithms \cite{Cohen97theaetg,lei1998}  within a dedicated constraint solver, producing predictable solutions. Both cannot handle thousand-sized FMs.  Recently, SPLCAT \cite{Johansen:2011:PRF:2050655.2050721}, used as a reference throughout this paper has been proposed. It also produces predictable solutions and handles larger FMs, but it does not scale to the Linux FM. An optimization of SPLCAT has been recently proposed \cite{johansen12}: it handles larger FMs than SPLCAT but is limited to $t=3$. While prioritization is induced in some approaches \cite{Oster:2010:AIP:1885639.1885658,Johansen:2011:PRF:2050655.2050721}, it is not explicit and thus not applicable directly. Logic was also used. Calvagna \etal explain how to deal with constraints in CIT \cite{Calvagna-logic-based} by encoding them in first order logic, offer various reductions algorithms to simplify them and use a model checker to solve them. Since this work was not related to FMs, it is difficult to assess its scalability. Hervieu \etal \cite{DBLP:conf/issre/HervieuBG11} also use reduction techniques in the aim of finding the minimal test suite in a Prolog-based implementation. However, this approach does not scale well to FMs of over 200 features, according to our experiments.                        

\subsubsection{Search-based Approaches}

Due to the computational complexity of $t$-wise testing of SPLs, using search-based heuristics is an option.  However,  we are only aware of two approaches \cite{DBLP:journals/ese/GarvinCD11,ensanevolutionary}. Garvin \etal \cite{DBLP:journals/ese/GarvinCD11} report on their experience applying and improving an extension to the AETG algorithm \cite{Cohen97theaetg} using simulated annealing. The simulated annealing approach incrementally populates a \emph{constrained covering array} \cite{Cohen2007} (which can be simply viewed as a table where lines represent products and columns features, like Table \ref{products_example})  by making some moves, i.e. changing the valuation of the features.  Each move is controlled by a SAT solver to ensure it is legal with respect to the FM constraints.  Moves are guided by a fitness function defined over the remaining pairs to be covered: the fewer pairs to be covered, the lower the probability to make a move.  As we have seen, using pairs coverage as a fitness function induces scalability issues which may be intractable for very large FMs or high $t$ values.  Similarly to ours,  Ensan \etal  devised a genetic algorithm approach to generate SPL test suites \cite{ensanevolutionary}.  They took a fine-grained perspective where each gene is a feature to be mutated and where crossover is applied, inducing possible invalid products which need to be removed.  Their fitness function indirectly measures coverage by evaluating the variability points to be bound and the constraints concerned by the features of a product. Rather, we adopt a coarse-grained approach which copes better with large FMs (\cite{ensanevolutionary} does not scale over 300 features) and does not produce invalid configurations (since a product is always replaced as a whole). As opposed to other approaches, Ensan \etal and our approaches yield partial $t$-wise coverage due to the choice of the fitness function. This, however, allows dealing more easily with time and cost constraints, looking for a ``good enough'' test suite.

\urlstyle{leo}
\balance
\section{Conclusion}\label{sConcl}


T-wise testing aims at finding bugs due to interactions amongst faulty features, which is particularly relevant in an SPL context. However, full $t$-wise testing is NP-complete and scalability an issue: no approach is able to handle high values of $t$ ($\geq$ 3) for large feature models in a reasonable amount of time (in days). Moreover, there is no suitable technique supporting the selection of only a fixed number of products, according to a limited budget. In this paper, we tackled these problem by proposing (a) approaches to prioritize products while maximizing the $t$-wise coverage and (b) a scalable and flexible search-based technique to generate products under budget and time constraints for large feature models. Both these techniques are computationally independent from $t$.

Our experiments, performed on 124 feature models for $t = 2$ to $t = 6$, show the feasibility and the scalability of our solutions. We managed to deal with the largest feature models available, such as the Linux kernel ($\approx$ 7,000 features, $\approx$ 200,000 constraints and $\approx$ 8.71E21 valid 6-sets) with up to 90.671\% of 6-wise coverage achieved with 1,000 products. Thus, by enabling a partial but scalable $t$-wise coverage and by introducing flexibility in the testing process, our approaches pave the way to a potentially $t$-unrestricted combinatorial interaction testing.

\bibliographystyle{IEEEtran}
\bibliography{IEEEabrv,bib}

\end{document}